# Thirty five classes of solutions of the quantum time-dependent two-state problem in terms of the general Heun functions

A.M. Ishkhanyan[1], T.A. Shahverdyan[1,2], and T.A. Ishkhanyan[1,2]

[1]Institute for Physical Research, NAS of Armenia, 0203 Ashtarak, Armenia
[1]Moscow Institute of Physics and Technology, Dolgoprudny, Moscow Region, 141700 Russia

We derive 35 five-parametric classes of the quantum time-dependent two-state models solvable in terms of the general Heun functions. Each of the classes is defined by a pair of generating functions the first of which is referred to as the amplitude- and the second one as the detuning-modulation function. The classes suggest numerous families of specific field configurations with different physical properties generated by appropriate choices of the transformation of the independent variable, real or complex. There are many families of models with constant detuning or constant amplitude, numerous classes of chirped pulses of controllable amplitude and/or detuning, families of models with double or multiple (periodic) crossings, periodic amplitude modulation field configurations, etc.

The detuning modulation function is the same for all the derived classes. This function involves four arbitrary parameters, that is, two more than the previously known hypergeometric classes. These parameters in general are complex and should be chosen so that the resultant detuning is real for the applied (arbitrary) complex-valued transformation of the independent variable. The generalization of the detuning modulation function to the four-parametric case is the most notable extension since many useful properties of the two-state models described by the Heun equation are due to namely the additional parameters involved in this function. Many of the derived amplitude modulation functions present different generalizations of the known hypergeometric models. In several cases the generalization is achieved by multiplying the amplitude modulation function of the corresponding prototype hypergeometric class by an extra factor including an additional parameter. Finally, many classes suggest amplitude modulation functions having forms not discussed before.

We present several families of constant-detuning field configurations generated by a real transformation of the independent variable. The members of these families are symmetric or asymmetric two-peak finite-area pulses with controllable distance between the peaks and controllable amplitude of each of the peaks. We show that the edge shapes, the distance between the peaks as well as the amplitude of the peaks are controlled almost independently, by different parameters. We identify the parameters controlling each of the mentioned features and discuss other basic properties of pulse shapes. We show that the pulse edges may become step-wise functions and determine the positions of the limiting vertical-wall edges. We show that the pulse width is controlled by only two of the involved parameters. For some values of these parameters the pulse width diverges and for some other values the pulses become infinitely narrow. We show that the effect of the two mentioned parameters is almost similar, that is, both parameters are able to independently produce pulses of almost the same shape and width. We determine the conditions for generation of pulses of almost indistinguishable shape and width, and present several such examples.

Finally, we present a constant-amplitude periodic level-crossing model and several families of constant-detuning field configurations generated by complex transformations of the independent variable.

# 1. Introduction

The two-state approximation is a widely known key approach intensively applied to study quantum non-adiabatic transitions in a number of important physical systems for many decades, already starting from early years of quantum mechanics [1-4]. This is a relatively simple, yet, useful and powerful approximation used to gather a basic insight of physical processes, to reveal the qualitative characteristics and underlying mechanisms [1-45]. Besides, numerous are the cases when the approximation provides also highly accurate quantitative description. The approximation basically assumes that the evolution of a physical system due to an interaction effectively consists in mixing of just two of quantum levels of the system; the change of others being negligible because of specific conditions considered.

In the canonical Landau-Zener-Majorana-Stückelberg formulation [1-4] the time-dependent version of the semi-classical quantum two-state problem is written as a system of coupled first-order differential equations for probability amplitudes of two states of a quantum system driven by an external field (time-dependent Schrödinger equations) [1-9]. The field is characterized by two real functions, the amplitude and the phase, composing a complex-valued function of a real argument, time. The analytic models of the time-dependent two-state problem of this form have been applied for studying of a number of important physical phenomena in many branches of contemporary physics, chemistry and engineering ranging from the theory of light-matter interaction, non-linear optics, surface physics to nanophysics, neutrino oscillations, cosmology, nuclear and chemical reactions, etc. The occurrences of the problem are too numerous to be mentioned in detail here. Apart from the earlier theory of magnetic resonance [1-4], some representative examples include the theory of atomic and molecular collisions [5,6], the general theory of non-adiabatic transitions [7], coherent atomic excitation [8] and controlling diverse quantum structures using laser radiation [9], laser cooling and trapping [10] using mechanical action of light [11], atom optics [12], theory of chemical reactions [13-15], quantum mechanical tunneling effects in inorganic, organic, and complex systems in biology [16-18], quantum phase transitions [19-22], quantum information [23,24], Bose-Einstein condensation [25-27], atom lasers [28-29], cold atom association in degenerate quantum gases [30-34], nanophysics [35-38], neutrino oscillations [39-41], etc.

The analytic models of the two-, three- and generally few-state problems developed in the past make use the (generalized) hypergeometric functions or their particular cases (see, e.g., [8-9,42-66] and references therein). In finding the solutions in terms of the Gauss hypergeometric and the Kummer confluent hypergeometric functions [42-55] as well as in terms of the Clausen [56-64] and Goursat [65,66] generalized hypergeometric functions [67-68], it has been recognized that the set of all solvable cases is divided into independent classes each containing infinite number of members generated by a corresponding basic integrable model [48-66]. The basic solvable models are defined by a pair of functions the first of which is referred to as the amplitude- and the second one as the detuning-modulation function. The actual field-configuration, that is the pair of the real functions defining the real physical field (e.g., the Rabi frequency and the frequency detuning as far as the quantum optical terminology is used) by means of application of a (generally complex-valued) transformation of the independent variable [52-55].

In the present paper we apply this general property of the solvable models of the time-dependent Schrödinger equations to derive solutions of the two-state problem in terms of the general Heun function [69]. This function is the solution of a second order linear differential equation having four regular singular points in the complex $z$-plane. This equation, the general Heun equation, directly generalizes the Gauss hypergeometric equation [70-72], and for this reason, it is routinely encountered in contemporary physics and mathematics research. It contains a large number of important special cases, in particular, the Mathieu and Lamé equations are the most studied ones which are of considerable importance in mathematical physics. The occurrences of the general Heun equation and its four confluent cases in classical and non-classical physics cover such fields as the rheology, diffusion, heat and mass transfer, magneto-hydrodynamics and wave mechanics, surface physics, polymer physics, condensed state physics, particle physics, general gravity, astronomy, cosmology, and many others (see, e.g., [73] and references therein). Due to extremely rich structure of the solutions of these equations the analytic developments based on them promise interesting and important new applications in many branches of contemporary physics and mathematics, and the special functions emerged from the solutions of these equations are supposed to gradually become a part of the next generation of the set of standard tools of mathematical physics.

We have previously discussed the reduction of the two-state problem to the confluent Heun equation and had found fifteen four-parametric classes of solvable models [74]. Below we show that the general Heun equation suggests much wider set of choices. Namely, we derive 35 five-parametric classes of solvable models which generalize all previously known integrable cases to more general classes and suggest many new families of solvable field configurations. The most notable feature of this generalization is due to the extension of the *detuning modulation* function to a four-parametric family suggesting many useful features. Besides, some of the derived *amplitude modulation* functions present different types of generalizations of the known hypergeometric ones, while many other classes suggest amplitude modulation functions having forms not discussed before. Finally, the derived classes also generalize the very few cases when the problem was treated using the general Heun equation [55,75,76].

Numerous specific field configurations with different physical properties can be generated by appropriate transformation of the independent variable, real or complex. Below we present several examples corresponding to the *constant detuning* case generated by *real* functions $z(t)$. Other such constant detuning configurations can be suggested by complex transformation, e.g., having the form $z = (1 + i y(t))/2$. Actually there are many other possible field configurations. For instance, there are numerous models with constant amplitude (in contrast to the case of constant detuning pulses), a variety of generally asymmetric chirped pulses of controllable amplitude and frequency detuning variation, symmetric or asymmetric models with double or multiple (periodic) crossings of the resonance, double or multiple level-glancing models, periodic amplitude modulation and bi-chromatic field configurations, etc. The examples are too many. Here we present only some of the families of constant detuning pulses generated by real functions $z(t)$ leaving the discussion of other possible field configurations for a separate consideration.

Discussing the constant-detuning field configurations we show that for 10 classes the members of such families are symmetric or asymmetric two-peak finite-area pulses with controllable distance between the peaks and controllable amplitude of each of the peaks. Notably, the distance between the peaks, the amplitude of the peaks as well as the shapes of the pulse edges are controlled almost independently, by different input parameters. We identify the parameters standing for each of these features and discuss the basic properties of pulse shapes. We mention some examples of field configurations of particular physical interest and discuss their general properties. In particular, we show that the pulse edges may be step-wise and determine the positions of the limiting vertical-wall edges. We show that the pulse width is controlled by only two of the involved parameters. However, we show that the effect of these parameters is rather similar, that is, both parameters are able to independently produce pulses almost of the same shape and width. We determine the conditions for generation of pulses of almost indistinguishable shape and width, and present a number of such examples. Finally, we show that pulses with arbitrary width can be produced. The cases of infinitely wide and infinitely narrow pulses are identified explicitly.

## 2. Reduction of the two-state problem to the Heun equation

The semiclassical time-dependent two-state problem is written as a system of coupled first-order differential equations for probability amplitudes $a_1(t)$ and $a_2(t)$ of given two states of a quantum system driven by an external field with amplitude modulation $U(t) > 0$, and phase modulation $\delta(t)$:

$$i\frac{da_1}{dt} = Ue^{-i\delta}a_2, \quad i\frac{da_2}{dt} = Ue^{+i\delta}a_1. \tag{1}$$

This system is equivalent to the following linear second-order ordinary differential equation:

$$a_{2tt}(t) + \left(-i\delta_t - \frac{U_t}{U}\right)a_{2t} + U^2 a_2 = 0, \tag{2}$$

where and hereafter the alphabetical index denotes differentiation with respect to corresponding variable. A useful property of solvable models of the two-state problem is that if the function $a_2^*(z)$ is a solution of this equation rewritten for an auxiliary argument $z$ for some functions $U^*(z)$, $\delta^*(z)$ then the function $a_2(t) = a_2^*(z(t))$ is the solution of Eq. (2) for the field-configuration defined as

$$U(t) = U^*(z)\frac{dz}{dt}, \quad \delta_t(t) = \delta_z^*(z)\frac{dz}{dt} \tag{3}$$

for arbitrary complex-valued function $z(t)$ [52-55]. We refer to the functions $U^*(z)$ and $\delta_z^*(z)$ as the amplitude- and detuning-modulation functions, respectively, and to the pair $\{U^*, \delta_z^*\}$ as basic integrable model.

Transformation of variables $z = z(t)$, $a_2 = \varphi(z)u(z)$ together with (3) reduces equation (2) to the following equation for the new dependent variable $u(z)$:

$$u_{zz} + \left(2\frac{\varphi_z}{\varphi} - i\delta_z^* - \frac{U_z^*}{U^*}\right)u_z + \left(\frac{\varphi_{zz}}{\varphi} + \left(-i\delta_z^* - \frac{U_z^*}{U^*}\right)\frac{\varphi_z}{\varphi} + U^{*2}\right)u = 0. \tag{4}$$

This equation is the general Heun equation [69]

$$u_{zz} + \left(\frac{\gamma}{z} + \frac{\delta}{z-1} + \frac{\varepsilon}{z-a}\right)u_z + \frac{\alpha\beta z - q}{z(z-1)(z-a)}u = 0, \tag{5}$$

where $1 + \alpha + \beta = \gamma + \delta + \varepsilon$, if

$$\frac{\gamma}{z} + \frac{\delta}{z-1} + \frac{\varepsilon}{z-a} = 2\frac{\varphi_z}{\varphi} - i\delta_z^* - \frac{U_z^*}{U^*} \tag{6}$$

and

$$\frac{\alpha\beta z - q}{z(z-1)(z-a)} = \frac{\varphi_{zz}}{\varphi} + \left(-i\delta_z^* - \frac{U_z^*}{U^*}\right)\frac{\varphi_z}{\varphi} + U^{*2}. \tag{7}$$

Searching for solutions of equations (6), (7) in the form

$$\varphi = z^{\alpha_1}(z-1)^{\alpha_2}(z-a)^{\alpha_3}, \qquad (8)$$

$$U^* = U_0^* z^{k_1}(z-1)^{k_2}(z-a)^{k_3}, \qquad (9)$$

$$\delta_z^* = \frac{\delta_1}{z} + \frac{\delta_2}{z-1} + \frac{\delta_3}{z-a}, \qquad (10)$$

where parameters $\alpha_{1,2,3}$, $U_0^*$, $k_{1,2,3}$ and $\delta_{1,2,3}$ are supposed to be constants, we multiply equation (7) by $z^2(z-1)^2(z-a)^2$ and note that then it follows from the obtained equation that the product $U_0^{*2} z^{2k_1+2}(z-1)^{2k_2+2}(z-a)^{2k_3+2}$ is a polynomial in $z$ of maximum fourth degree: $-1 \le k_{1,2,3} \cup k_1+k_2+k_3 \le -1$. This leads to 35 possible choices of $k_{1,2,3}$. These sets are shown on Fig. 1 by points in 3D-space of parameters $k_{1,2,3}$. The corresponding basic models $U^*(z)$ are explicitly presented in Tables 1-5. Note that since the parameters $a$, $U_0^*$ and $\delta_{1,2,3}$ remain arbitrary, all the derived classes are 5-parametric.

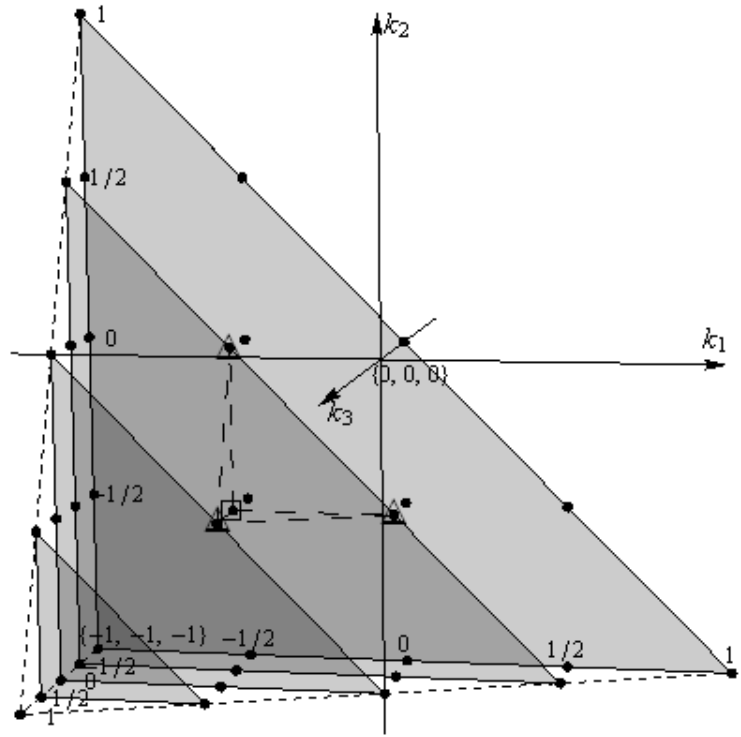

Fig. 1. Thirty five points in the parameter space $\{k_1, k_2, k_3\}$ defining the 35 classes of 5-parametric models of the two-state problem solvable in terms of the general Heun equation. The four cases for which $\varphi = 1$ are indicated by three triangles and a square.

## 3. Thirty five basic integrable models

As it is seen, we have 15 basic models with $k_3 = -1$ (Table 1) - this is the richest subset of classes. The physical field-configurations $\{U(t), \delta(t)\}$ generated by the six models closest to the lower left corner generalize the six well known hypergeometric models (compare with the classes of Table 3), widely discussed in the past by many authors (see, e.g., [8,9], [45-55] and references therein), by the extra factor $1/(z-a)$. Note that along with this extension, an additional generalization comes from extra term $\delta_3/(z-a)$ in Eq. (10). Other models suggest different extensions of $U^*(z)$. Three-parametric families of field configurations belonging to the classes $\{k_1, k_2, k_3\} = \{-1/2, k_2, -1\}$ with $k_2 = -1/2, 0, 1/2$, for which the solution of the two-state problem is written in terms of the Gauss hypergeometric functions $_2F_1$, were presented in [55].

| $k_2$ \ $k_1$ | $-1$ | $-1/2$ | $0$ | $1/2$ | $1$ |
|---|---|---|---|---|---|
| $1$ | $\dfrac{z-1}{z(z-a)}$ | | | | $k_3 = -1$ |
| $1/2$ | $\dfrac{\sqrt{z-1}}{z(z-a)}$ | $\dfrac{\sqrt{z-1}}{\sqrt{z}(z-a)}$ | | | |
| $0$ | $\dfrac{1}{z(z-a)}$ | $\dfrac{1}{\sqrt{z}(z-a)}$ | $\dfrac{1}{z-a}$ | | |
| $-1/2$ | $\dfrac{1}{z\sqrt{z-1}(z-a)}$ | $\dfrac{1}{\sqrt{z(z-1)}(z-a)}$ | $\dfrac{1}{\sqrt{z-1}(z-a)}$ | $\dfrac{\sqrt{z}}{\sqrt{z-1}(z-a)}$ | |
| $-1$ | $\dfrac{1}{z(z-1)(z-a)}$ | $\dfrac{1}{\sqrt{z}(z-1)(z-a)}$ | $\dfrac{1}{(z-1)(z-a)}$ | $\dfrac{\sqrt{z}}{(z-1)(z-a)}$ | $\dfrac{z}{(z-1)(z-a)}$ |

Table 1. Fifteen basic models $U^*/U_0^*$ with $k_3 = -1$. The field-configurations $\{U(t), \delta(t)\}$ generated by the six models closest to the lower left corner generalize the well known six hypergeometric models [53-55] (see Table 3) by the extra factor $1/(z-a)$ (note that along with this extension, additional generalization comes from the extra term $\delta_3/(z-a)$ in Eq. (10)). Other models suggest different extensions of $U^*(z)$. Three-parametric subfamilies of field configurations belonging to the classes $\{k_1, k_2, k_3\} = \{-1/2, k_2, -1\}$ with $k_2 = -1/2, 0, 1/2$, for which the solution of the two-state problem is written in terms of the Gauss hypergeometric functions $_2F_1$, were presented in [55].

The next 10 basic models correspond to the choice $k_3 = -1/2$ (Table 2). Here again, the six models closest to the lower left corner generalize the hypergeometric models [53-55] (compare with the classes of Table 3), this time, by the extra factor $1/\sqrt{z-a}$ (note that along with this extension, here also a further generalization comes from the extra term $\delta_3/(z-a)$ in Eq. (10)), while the four diagonal models are of different structure.

Six basic models correspond to $k_3 = 0$ (Table 3). Though these models are exactly of the same form as the ones solvable in terms of the Gauss hypergeometric equation (see [53-55]), however, due to the additional term $\delta_3/(z-a)$ in equation (10) for the detuning modulation, the final physical field-configurations $\{U(t),\delta(t)\}$ generated by these models are more general and include the hypergeometric models as particular cases.

Finally, 3 basic models correspond to $k_3 = 1/2$ (Table 4), and there is only one model in the case $k_3 = 1$: $\{k_1, k_2, k_3\} = \{-1, -1, +1\}$ (Table 5). The three basic models of Table 4 generalize the hypergeometric basic models by the extra factor $\sqrt{z-a}$, and the model of Table 5 generalizes the corresponding hypergeometric model $U^*/U_0^* = 1/(z(z-1))$ by the factor $z-a$. As in the case of all other general Heun models, further generalization for all these classes consists in the term $\delta_3/(z-a)$ in Eq. (10).

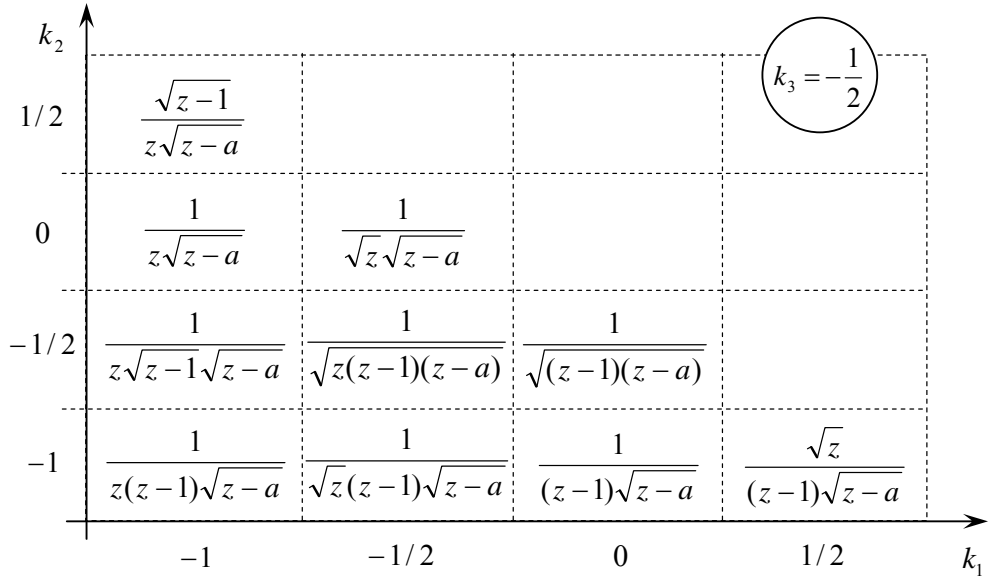

Table 2. Ten basic models $U^*/U_0^*$ corresponding to $k_3 = -1/2$. The six models closest to the lower left corner generalize the hypergeometric models [53-55 (see Table 3) by the extra factor $1/\sqrt{z-a}$ (note that along with this extension, further generalization comes from the extra term $\delta_3/(z-a)$ in Eq. (10)), while the four diagonal models are of different structure.

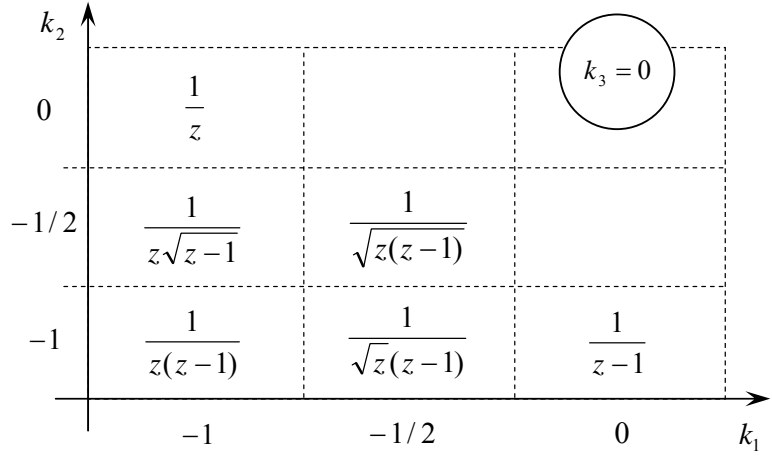

Table 3. Six basic models $U^*/U_0^*$ corresponding to $k_3 = 0$. Though these models are exactly of the same form as the ones for the Gauss hypergeometric equation (see [55]), however, due to the additional term $\delta_3/(z-a)$ in the equation (10) for the detuning modulation, the final field-configurations $\{U(t),\delta(t)\}$ generated by these models are more general and include the hypergeometric models as particular cases.

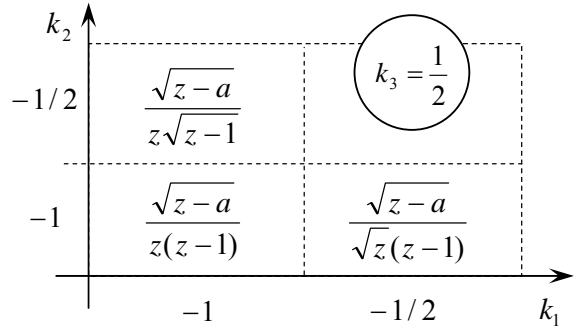

Table 4. Three basic models $U^*/U_0^*$ corresponding to $k_3 = 1/2$. All the three generalize the hypergeometric models by the extra factor $\sqrt{z-a}$ as well as, as in the case of all other Heun models, further generalization comes from the extra term $\delta_3/(z-a)$ in Eq. (10).

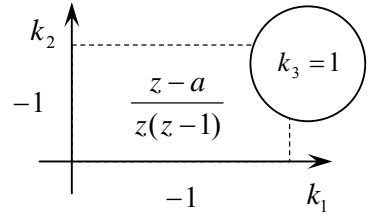

Table 5. The only basic model $U^*/U_0^*$ corresponding to $k_3 = 1$. The generalization of the corresponding hypergeometric model $U^*/U_0^* = 1/(z(z-1))$ consists in the extra factor $z-a$ and, furthermore, in the extra term $\delta_3/(z-a)$ in Eq. (10).

## 4. Series solutions of the general Heun equation

According to the property (3) of integrable models and Eqs. (9), (10), the physical field-configuration, i.e., the Rabi frequency and the frequency detuning of the laser radiation if quantum optical terminology is applied, are given as

$$U(t) = U_0^* z^{k_1} (z-1)^{k_2} (z-a)^{k_3} \frac{dz}{dt}, \qquad (11)$$

$$\delta_t(t) = \left( \frac{\delta_1}{z} + \frac{\delta_2}{z-1} + \frac{\delta_3}{z-a} \right) \frac{dz}{dt}. \qquad (12)$$

Note that the parameters $a$, $U_0^*$, $\delta_{1,2,3}$ are in general *complex* constants which should be chosen so that the functions $U(t)$ and $\delta(t)$ are real for given *complex-valued* function $z(t)$.

The solution of the initial two-state problem (1) is explicitly written as

$$a_2 = C_0 z^{\alpha_1} (z-1)^{\alpha_2} (z-a)^{\alpha_3} H(a,q;\alpha,\beta;\gamma,\delta;z), \quad C_0 = \text{const}, \qquad (13)$$

where $H(a,q;\alpha,\beta;\gamma,\delta;z)$ is the Heun function satisfying Eq. (5) and $\alpha_{1,2,3}$ are defined by the following quadratic equations:

$$\begin{aligned}
\alpha_1(\alpha_1 + 1 - \gamma) &= \left( z^2 U^{*2} \right)\Big|_{z=0}, \\
\alpha_2(\alpha_2 + 1 - \delta) &= \left( (z-1)^2 U^{*2} \right)\Big|_{z=1}, \\
\alpha_3(\alpha_3 + 1 - \varepsilon) &= \left( (z-a)^2 U^{*2} \right)\Big|_{z=a}.
\end{aligned} \qquad (14)$$

The Heun function's parameters $\gamma, \delta, \varepsilon$ and $q$ are given through the equations

$$\gamma = 2\alpha_1 - i\delta_1 - k_1, \ \delta = 2\alpha_2 - i\delta_2 - k_2, \ \varepsilon = 2\alpha_3 - i\delta_3 - k_3, \qquad (15)$$

$$q = (\gamma - 2\alpha_1)(\alpha_3 + a\alpha_2) + (a\delta + \varepsilon)\alpha_1 - a\left( zU^{*2} - \alpha_1(\alpha_1 + 1 - \gamma)/z \right)\Big|_{z=0}, \qquad (16)$$

and $\alpha, \beta$ are determined from the Fuchsian condition $1 + \alpha + \beta = \gamma + \delta + \varepsilon$ together with the following equation

$$\begin{aligned}
\alpha\beta &= q + \varepsilon\alpha_2 + \alpha_3(\delta - 2\alpha_2) - (1-a)(2\alpha_1\alpha_2 - \delta\alpha_1 - \gamma\alpha_2) \\
&+ (1-a)\left( (z-1)U^{*2} - \alpha_2(\alpha_2 + 1 - \delta)/(z-1) \right)\Big|_{z=1}.
\end{aligned} \qquad (17)$$

A solution of the general Heun equation can be constructed as a Frobenius power-series expansion:

$$H(a,q;\alpha,\beta;\gamma,\delta;z) = z^\mu \sum_n c_n z^n. \qquad (18)$$

The coefficients of this expansion obey the three-term recurrence relation [70-72]

$$R_n c_n + Q_{n-1} c_{n-1} + P_{n-2} c_{n-2} = 0, \tag{19}$$

where
$$R_n = a(\mu + n)(\mu + n - 1 + \gamma), \tag{20}$$

$$Q_n = -q - (\mu + n)\big((\mu + n - 1 + \gamma + \delta + \varepsilon)(1 + a) - a\varepsilon - \delta\big), \tag{21}$$

$$P_n = (\mu + n + \alpha)(\mu + n + \beta). \tag{22}$$

For left-hand side termination of the series at $n = 0$ (i.e., $c_0 = 1$ and $c_{-2} = c_{-1} = 0$), should be $R_0 = 0$, so that one should choose

$$\mu = 0 \quad \text{or} \quad \mu = 1 - \gamma. \tag{23}$$

The convergence radius of this series is $\min\{|a|, 1\}$. The series is right-hand side terminated at some $n = N$ if $c_N \neq 0$ and $c_{N+1} = c_{N+2} = 0$. Hence, should be $P_N = 0$, that is

$$\mu + \alpha = -N \quad \text{or} \quad \mu + \beta = -N, \tag{24}$$

and
$$Q_N c_N + P_{N-1} c_{N-1} = 0. \tag{25}$$

The last equation is a polynomial equation of the order $N + 1$ for the accessory (spectral) parameter $q$ having in general $N + 1$ solutions.

Alternatively, instead of powers, one may apply other expansion functions. For instance, several expansions in terms of the Gauss hypergeometric functions [77-84] and incomplete beta functions [85-88] are known. Using the properties of the derivatives of the solutions of the Heun equation [88-90], expansions in terms of higher transcendental functions, e.g., the Goursat hypergeometric functions or Appell hypergeometric function of two variables, can be constructed [88,91]. The region of convergence in these cases may be other than a circle. Besides, the expansions apply to different combinations of involved parameters, so that they may be useful for applications in different physical situations. We would like to mention here an expansion in terms of the Gauss hypergeometric functions that is convenient for derivation of closed form solutions [82]:

$$u = \sum_n c_n u_n = \sum_n c_n \cdot {}_2F_1(\alpha, \beta; \gamma_0 - n; z), \tag{26}$$

where the coefficients of the expansion obey the following three-term recurrence relation:

$$R_n c_n + Q_{n-1} c_{n-1} + P_{n-2} c_{n-2} = 0 \tag{27}$$

with
$$R_n = \frac{a}{\gamma_0 - n}(\gamma - \gamma_0 + n)(\alpha - \gamma_0 + n)(\beta - \gamma_0 + n), \tag{28}$$

$$Q_n = (1 - a)(\varepsilon + \gamma - \gamma_0 + n)(\gamma_0 - n - 1) + a(\gamma - \gamma_0 + n)(\alpha + \beta - \gamma_0 + n) + \alpha\beta a - q, \tag{29}$$

$$P_n = (a - 1)(\varepsilon + \gamma - \gamma_0 + n)(\gamma_0 - n - 1). \tag{30}$$

The series is terminated from the left-hand side at $n=0$ ($c_0 = 1$, $c_{-2} = c_{-1} = 0$) if

$$\gamma_0 = \gamma \text{ or } \alpha \text{ or } \beta. \tag{31}$$

Furthermore, the series is right-side terminated at some $n = N$ ($c_N \neq 0$, $c_{N+1} = c_{N+2} = 0$) if $P_N = 0$. Since $\gamma_0$ cannot be a positive integer, we get $\gamma_0 = \varepsilon + \gamma + N$ or, equivalently,

$$\varepsilon, \varepsilon + \gamma - \alpha \text{ or } \varepsilon + \gamma - \beta = -N \tag{32}$$

if $\gamma_0 = \gamma$ or $\alpha$ or $\beta$, respectively. The second condition for termination of the series, $Q_N c_N + P_{N-1} c_{N-1} = 0$, is polynomial equation of the order $N+1$ for the accessory (spectral) parameter $q$ having in general $N+1$ solutions.

## 5. Constant detuning models: real $z(t)$

Eqs. (11) and (12) provide a rich variety of field configurations generated by different choices of the transformation $z(t)$. Here we restrict ourselves by discussion of the basic case of *constant detuning* field configurations generated by *real* choices of this transformation.

Thus, let the detuning be constant: $\delta_t(t) = \Delta = \text{const}$. For a real $z(t)$, in order to ensure that the above power series expansion is applicable, let $z \in (0,1)$ and $a > 1$. Then, the families of constant-detuning field configurations are defined parametrically as:

$$t - t_0 = \ln\left(z^{\delta_1/\Delta}(1-z)^{\delta_2/\Delta}(a-z)^{\delta_3/\Delta}\right), \tag{33}$$

$$U(t) = \Delta \frac{U_0^* z^{k_1+1}(z-1)^{k_2+1}(z-a)^{k_3+1}}{(\delta_1 + \delta_2 + \delta_3)z^2 - (\delta_1 + a\delta_1 + a\delta_2 + \delta_3)z + a\delta_1}. \tag{34}$$

To specify the integration constant $t_0$, which only defines a nonessential shift in time, we may demand $z(t=0) = 1/2$, hence, $t_0 = \ln\left(2^{(\delta_1+\delta_2+\delta_3)/\Delta}(2a-1)^{-\delta_3/\Delta}\right)$.

The derived families of pulses include both symmetric and asymmetric members. The pulses may or may not vanish at $t \to \pm\infty$. Among all the families only 10 provide pulses for which the pulse area is finite, i.e. the pulses vanish at infinity. These are the families for which $k_{1,2} \neq -1$. The members of the latter families in general are two-peak symmetric or asymmetric pulses with controllable distance between the peaks and controllable amplitude of each of the peaks. We will see that the distance between the peaks as well as the amplitude of the peaks is mainly controlled by the parameters $a$ and $\delta_3$, while the sharpness of the pulse edges is mostly controlled by $\delta_1$ and $\delta_2$. Some representative examples of pulse shapes for different families are shown in Figs. 2-8.

Discussing the general properties of the pulses, it is natural to expect that new properties, compared to the known 6 hypergeometric classes [53-55], should be due to the parameters $a$ and $\delta_3$ which are not present in the hypergeometric case. To reveal the role of these parameters, we note that the denominator in the Eq. (34) for $U(t)$ for $z \to 0$ and $z \to 1$ tends to $a\delta_1$ and $(1-a)\delta_2$, respectively. Hence, within a fixed class defined by a chosen set of $k_{1,2,3}$, for fixed $a$ and $\delta_3$ the behavior of $U(t)$ for $t \to -\infty$ and $t \to +\infty$ is controlled by the parameters $\delta_1$ and $\delta_2$, respectively. For the families of pulses, for which $U(t)$ vanishes at $t \to \pm\infty$, at $\delta_1 \to +0$ the left edge of the pulse becomes a step-wise function and for $\delta_2 \to -0$ the same happens with the right edge (we consider the case $z'(t) > 0$, see the discussion below). In the simultaneous limit $\delta_{1,2} \to 0$ we have a pulse with vertical walls. These observations are demonstrated in Fig. 2 where the normalized ($U_{max} = 1$) pulse shapes for small $\delta_1$ (left figure) or small $\delta_2$ (right figure) are shown for the class $k_{1,2,3} = \{0,0,-1\}$.

As it is seen, for the class $k_{1,2,3} = \{0,0,-1\}$ the simultaneous limit $\delta_{1,2} \to 0$ produces a rectangular box-pulse. For other classes, of course, the pulse shapes are different, see for example Fig. 3. An important common feature, however, is that the width of the pulse is controlled by the parameters $a$ and $\delta_3$. This is demonstrated in Figs. 3 and 4. Note that the pulse width diverges as $a$ goes to the unity or as $\delta_3 \to -\infty$.

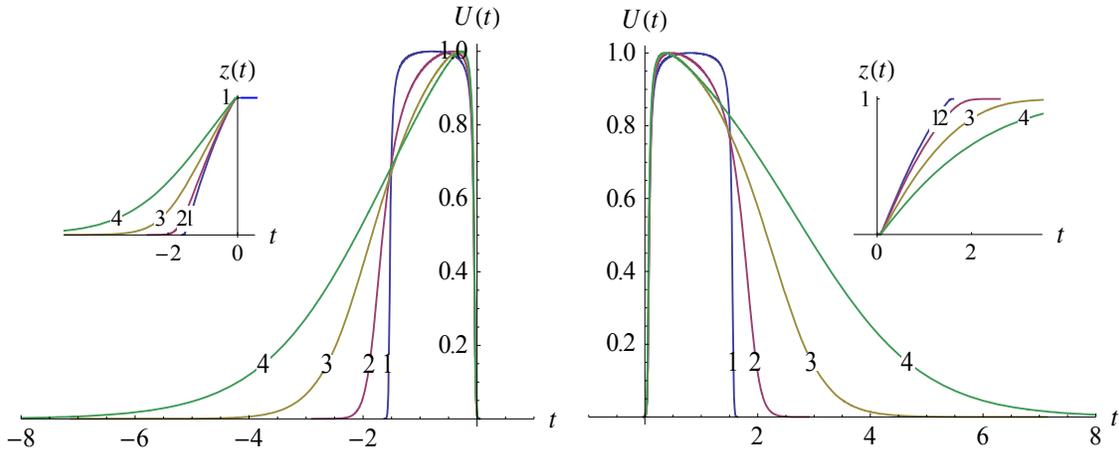

Fig. 2. Class $k_{1,2,3} = \{0,0,-1\}$. Constant detuning case, $\Delta = 1$. Normalized ($U_{max} = 1$) pulse shapes for $a = 2$, $\delta_3 = -2$. Graphs in the left figure correspond to $\delta_1 = 0.01; 0.1; 0.4; 1$ (curves 1,2,3,4, respectively) and small $\delta_2 = -0.01$. Right figure's graphs correspond to $\delta_2 = -0.01; -0.1; -0.4; -1$ (curves 1,2,3,4, respectively) and small $\delta_1 = 0.01$.

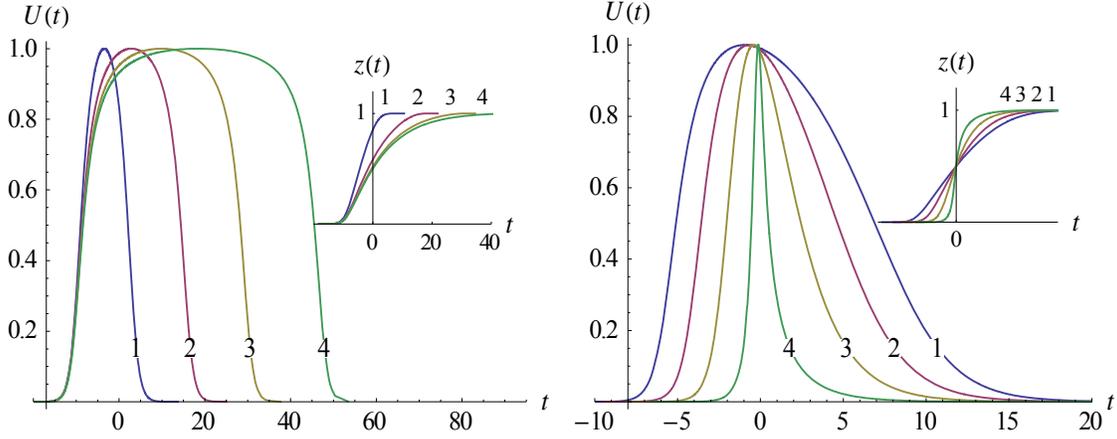

Fig. 3. Class $k_{1,2,3} = \{0, 0, -1\}$. Constant detuning case, $\Delta = 1$. Normalized ($U_{max} = 1$) pulse shapes for $\delta_1 = 1$, $\delta_2 = -1$. Graphs in the left figure correspond to $\delta_3 = -10$ and $a = 2; 1.2; 1.05; 1.01$ (curves 1,2,3,4, respectively). The same pulses are obtained by fixing, e.g., $a = 2$ and changing $\delta_3$: $\delta_3 = -10; -26.22; -44.87, -68.85$. The pulse width diverges as $a$ goes to the unity or as $\delta_3 \to -\infty$. Graphs in the right figure correspond to $a = 2$, $\delta_1 = 1/2$, $\delta_2 = -2$ and $\delta_3 = -10; -5; 0; 5$ (curves 1,2,3,4, respectively). Infinitely narrow pulse is achieved if $\delta_3 = 2\sqrt{2} + 9/2$.

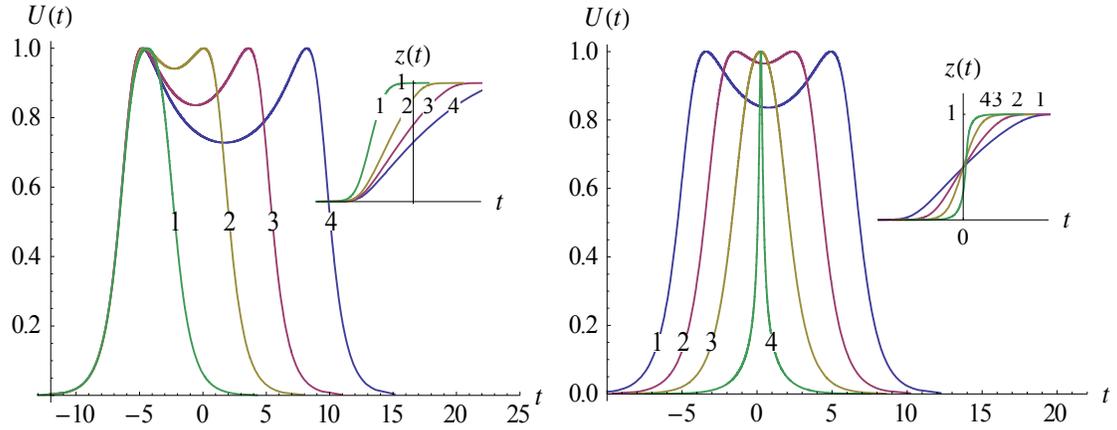

Fig. 4. Class $k_{1,2,3} = -1/2$. Constant detuning case, $\Delta = 1$. Normalized ($U_{max} = 1$) pulse shapes for $\delta_1 = 1/2$, $\delta_2 = -1/2$. Graphs in the left figure correspond to $\delta_3 = -10$ and $a = 10; 3; 2; 1.5$ (curves 1,2,3,4, respectively). The pulse width diverges as $a$ goes to the unity or as $\delta_3 \to -\infty$. The same pulses are produced by fixing an arbitrary chosen $a$ and varying $\delta_3$. For example, for $a = 2$ the result reads $\delta_3 = -1.52; -5.85; -10, -15.85$. On the right figure, curves 1,2,3,4 correspond to $a = 2$ and $\delta_3 = -10; -5; -1; 2.1$, respectively. Infinitely narrow pulse is achieved when $\delta_3 = 2\sqrt{2} + 3/2$.

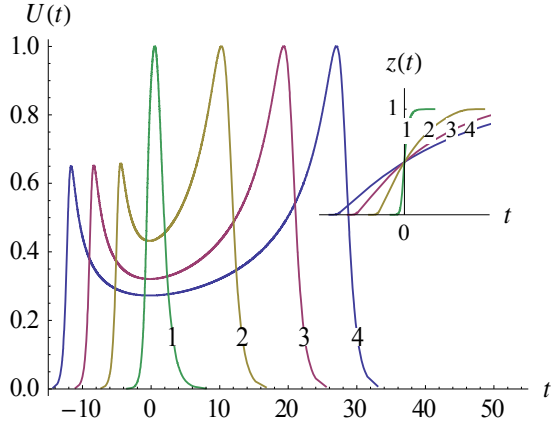

Fig. 5. Class $k_{1,2,3} = \{-1/2, -1/2, -1\}$.
Constant detuning case, $\Delta = 1$.
Normalized pulse shapes for $a = 1.2$,
$\delta_1 = 1/5$, $\delta_2 = -1/2$ and
$\delta_3 = 0, -7, -14, -20$ (curves 1,2,3,4).

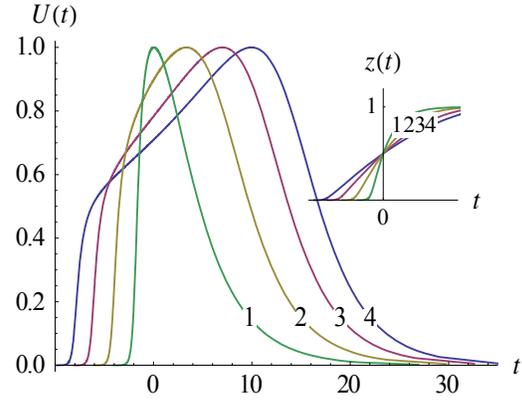

Fig. 6. Class $k_{1,2,3} = \{0, -1/2, -1\}$.
Constant detuning case, $\Delta = 1$.
Normalized pulse shapes for $a = 2$,
$\delta_1 = 1/5$, $\delta_2 = -2$ and
$\delta_3 = 0, -7, -14, -20$ (curves 1,2,3,4).

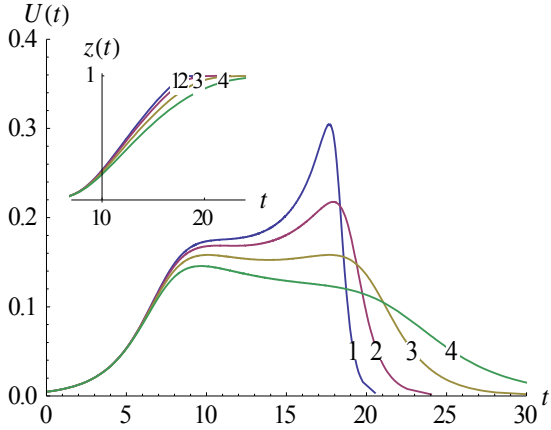

Fig. 7. Class $k_{1,2,3} = -1/2$. Constant
detuning case, $\Delta = 1$. Pulse shapes for
$U_0^* = -1$, $a = 2$, $\delta_1 = 1$, $\delta_3 = -10$. Curves
1,2,3,4 correspond to
$\delta_2 = -1/4, -1/2, -1, -1.8$, respectively.

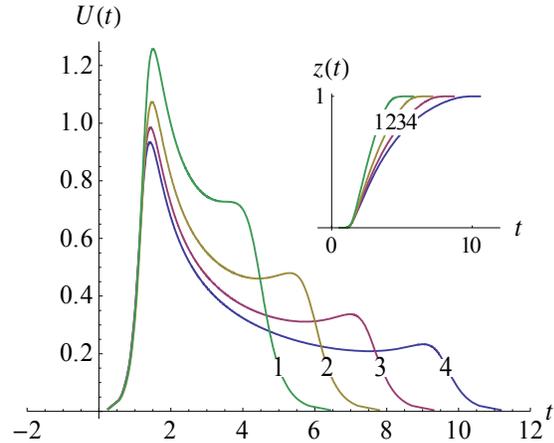

Fig. 8. Class $k_{1,2,3} = \{-1/2, -1/2, 0\}$.
Constant detuning case, $\Delta = 1$. Pulse
shapes for $U_0^* = i$, $\delta_1 = 1/10$, $\delta_2 = -1/5$,
$\delta_3 = -3$. Curves 1,2,3,4 correspond to
$a = 2.1, 1.5, 1.25, 1.12$, respectively.

Furthermore, it can be checked that the effect of the two parameters is rather similar, namely, in many cases both parameters are able to produce similar pulses with the same width. For instance, it can be shown that for $\delta_{1,2} \leq 1$ any two pairs $\{a, \delta_3\}$ and $\{a_0, \delta_{30}\}$ related by the equation $\delta_3 / \ln((a-1)/a) = \delta_{30} / \ln((a_0-1)/a_0)$ produce pulses of almost indistinguishable shapes. In the captions of Figs. 3 and 4 several such pairs are presented. Thus, the families (33), (34) allow variation of the edge shapes and pulse width almost independently, using different parameters.

## 6. Infinitely narrow pulses

Consider now whether infinitely narrow pulses are possible. First, we recall that Eq. (33) defines one-to-one mapping $t \leftrightarrow z$ if $z'(t)$ is sign-preserving. Hence, the polynomial

$$P(z) = \delta_1(z-1)(z-a) + \delta_2 z(z-a) + \delta_3 z(z-1), \tag{35}$$

the numerator of the term in brackets before $z'(t)$ in Eq. (12), should not change its sign on the interval $z \in (0,1)$. On the other hand, it follows from Eq. (34) that the pulse should diverge at some point $z_0$ belonging to this interval in order to become infinitely narrow after being normalized to $U_{\max} = 1$. This is achieved if the denominator of the ratio in the right-hand side of Eq. (34) vanishes at $z_0$. Since this denominator is the above polynomial $P(z)$, we conclude that $z_0$ is a root of the equation $P(z_0) = 0$. Furthermore, in order the polynomial to be sign-preserving, $z_0$ should be a multiple root of $P(z)$. Hence, the discriminant of the polynomial should be zero. Thus, we get the following conditions for the pulse parameters to produce infinitely narrow pulse: $D = 0$, $z_0 \in (0,1)$, where

$$D = (\delta_1 + a\delta_1 + a\delta_2 + \delta_3)^2 - 4a\delta_1(\delta_1 + \delta_2 + \delta_3), \tag{36}$$

$$z_0 = (\delta_1 + a\delta_1 + a\delta_2 + \delta_3)/(2(\delta_1 + \delta_2 + \delta_3)). \tag{37}$$

The equation $D = 0$ is a quadratic equation with respect to both $a$ and $\delta_3$. However, if considered as an equation for $a$, for some set of involved parameters it may have roots not belonging to the supposed variation interval of $a$: $a \in (1, +\infty)$. In contrary, no restriction is imposed on the variation range of $\delta_3$. This is a slight difference between the two considered parameters. Summarizing, we may state that the pulse width may go to zero or a non-zero minimum in both cases: when $a$ is fixed and $\delta_3$ is varied and vice versa. The limit of infinitely narrow pulse achieved via variation of $\delta_3$ is demonstrated in Fig. 3 (right figure) for asymmetric pulses and in Fig. 4 (right figure) for symmetric pulses. Here, the examples of the classes $k_{1,2,3} = \{0, 0, -1\}$ and $k_{1,2,3} = -1/2$ are presented, however, the above discussion is general and applies to all the families (33), (34) for which the pulse vanishes at $t \to \pm+\infty$.

## 7. Vertical edge pulses of controllable width

Next interesting point is how the pulse edge becomes a vertical wall at a limit $p \to p_0$ for an involved parameter $p$. It is understood that in order this to occur the derivative $U'(t)$ should diverge for a *fixed* time point $t_0$ if the limit $p \to p_0$ is considered. It can be checked that, in terms of $z$, $U'(t)$ is a polynomial in $z$ divided by $P(z)^3$, where $P(z)$ is the quadratic polynomial given by Eq. (35). Hence, $P(z(t_0))$ should go to zero at the limit $p \to p_0$. This means that $z(t_0)$ should tend to a *root* of the polynomial $P(z)$. However, we recall that $P(z)$ is sign-preserving on the interval $z \in (0,1)$ so that it cannot have real roots on this interval. Hence, the only possibility left is that the function $z(t_0)$ and a root of the polynomial $P(z)$, $z_1$ or $z_2$, as functions of the parameter $p$, should simultaneously tend to an endpoint of the segment $z \in [0,1]$. If $z_1 < 0$ and $z_2 > 1$, then for the left edge should hold $z_1(p \to p_0) = 0 = z(t_0, p \to p_0)$ and for the right edge should be $z(t_0, p \to p_0) = 1 = z_2(p \to p_0)$. With these observations, we consider the behavior of the pulse in the vicinity of the points $z = 0$ and $z = 1$.

For $z \to 0$, expanding $\ln(1-z)$ and $\ln(a-z)$ in the right-hand side of Eq. (33) in powers of $z$ and keeping only the constant term we get $t(z) = t_0 + (\delta_1 \ln(z) + \delta_3 \ln(a))/\Delta$ so that in this approximation $z(t) = \mathrm{Exp}((\Delta t - \Delta t_0 - \delta_3 \ln a)/\delta_1)$. This function, however, leads to a pulse which diverges at the time point $t_1 = t_0 + \delta_3 \ln(a)/\Delta$. Much better approximation is achieved using the next, linear-in-$z$, term in the expansion of mentioned two logarithms: $t(z) = t_0 + (\delta_1 \ln(z) + \delta_3 \ln(a) - (\delta_2 + \delta_3/a)z)/\Delta$. The solution of this equation is given by the Lambert $W$ function [92,93]:

$$z(t) = -\frac{a\delta_1}{a\delta_2 + \delta_3} W\left(-\frac{a\delta_2 + \delta_3}{\delta_1} e^{\frac{(t-t_0)\Delta - (\delta_1 + \delta_3)\ln a}{\delta_1}}\right). \tag{38}$$

This is a good approximation that leads to an accurate description of the pulse shape near its left edge for all the variation range of the involved parameters (see Fig. 9). For small $\delta_1$ it describes a jump with the width proportional to $\delta_1$. We then conclude, from this solution, that the pulse edge becomes a vertical wall at the limit $\delta_1 \to 0$ and the position of the wall is

$$t_1 = t_0 + \delta_3 \ln a / \Delta . \tag{39}$$

In the similar way, expanding now the logarithms $\ln(z)$ and $\ln(a-z)$ in the right-hand side of Eq. (33) in powers of $1-z$ and taking the constant and linear-in-$z$ terms we get an accurate description for the right edge in terms of another Lambert $W$ function. From this solution we get that this time the wall is formed at the limit $\delta_2 \to 0$ and its position is

$$t_2 = t_0 + \delta_3 \ln(a-1) / \Delta . \tag{40}$$

Thus, in the simultaneous limit $\delta_{1,2} \to 0$ the pulse width is given as

$$d = t_2 - t_1 = \delta_3 \ln((a-1)/a) / \Delta . \tag{41}$$

This formula shows that the pulse width is mainly controlled by the parameters $\{\delta_3, a\}$ and explains the above mentioned condition of having almost indistinguishable pulses of the same width using different pairs $\{\delta_3, a\}$. The comparison of the pulse shape provided by the approximation (38) as well as the divergent solution generated by the simple exponential approximation for $z(t)$ with the exact pulse shape is shown on Fig. 9. It is seen, that in the limit $\delta_1 \to 0$ the divergent solution accurately determines the position of the wall.

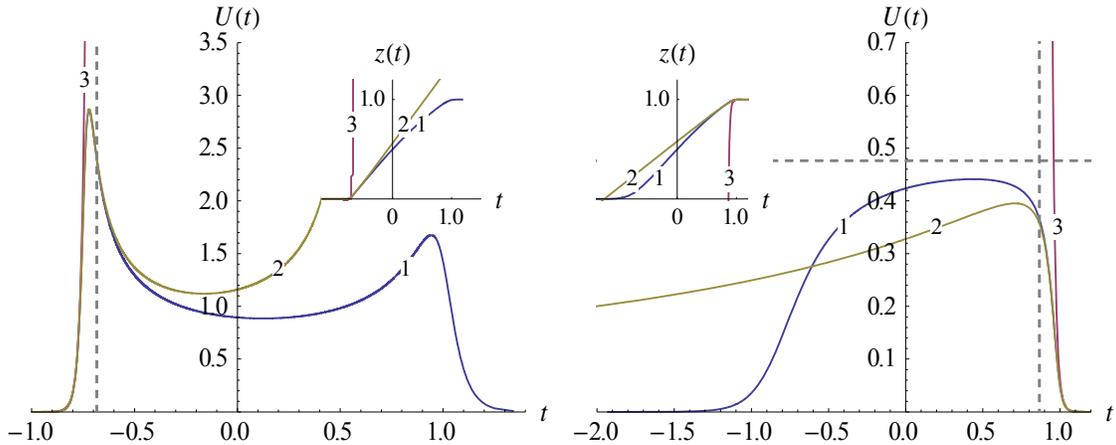

Fig. 9. Constant detuning case, $\Delta = 1$. Exact pulse shapes (curves 1) compared with those given by approximation (38) (curves 2). The divergent solutions generated by the simple exponential approximation for $z(t)$ (curves 3) accurately determine the limiting positions of the left and right edges shown by vertical dashed lines. Left graph: class $k_{1,2,3} = -1/2$, $U_0^* = -1$, $a = 2.5$, $\delta_1 = 0.01$, $\delta_2 = -0.03$, $\delta_3 = -3$. Right graph: class $k_{1,2,3} = \{0,0,-1\}$, $U_0^* = -1$, $a = 2$, $\delta_1 = 0.1$, $\delta_2 = -0.02$, $\delta_3 = -2$. The horizontal dashed line defines the height of the limiting box pulse.

## 8. Complex-valued transformation $z(t)$

A different set of pulses is generated by Eqs. (11),(12) if a *complex-valued* transformation $z = x(t) + i y(t)$ is applied. Consider, for instance, constant-detuning field configurations achieved by the substitution $z = (1 + i y(t))/2$. It follows from Eq. (11) that real amplitude-modulation functions are generated by this transformation if $k_1 = k_2$. Only 9 classes out of 35 satisfy this condition: $k_{1,2} = -1$, $k_3 = \{-1, -1/2, 0, 1/2, 1\}$; $k_{1,2} = -1/2$, $k_3 = \{-1, -1/2, 0\}$; and $k_{1,2} = 0$, $k_3 = -1$.

The corresponding pulse shapes are given parametrically as

$$t = \lambda_1 \ln(1 + y^2) + 2\lambda_2 \arctan(y) + \lambda_3 \ln\left(\frac{a_0 - y}{a_0}\right), \quad (42)$$

$$U(t) = \frac{U_0 (1 + y^2)^{k_1 + 1} (y - a_0)^{k_3 + 1}}{2(y - a_0)(\lambda_2 + \lambda_1 y) + \lambda_3 (1 + y^2)}, \quad (43)$$

where without loss of generality we have supposed $y(0) = 0$ and introduced real parameters $a_0$, $\lambda_{1,2,3}$ and $U_0$: $a = (1 + i a_0)/2$, $\delta_{1,2}/\Delta = \lambda_1 \mp i\lambda_2$, $\delta_3/\Delta = \lambda_3$, $U_0^* = (-2i)^{1+2k_1+k_3} U_0$. It can be seen from Eq (42) that if $a_0 > 0$, then $y(t) \in (-\infty, a_0)$ and if $a_0 < 0$, then $y(t) \in (a_0, +\infty)$. For definiteness, we consider the case $a_0 < 0$. Then, at an appropriate choice of the remaining parameters, Eqs. (42)-(43) define asymmetric pulse shapes shown in Figs. 10,11. Note that the pulses of the subfamilies $k_{1,2} = -1$, $k_3 = \{-1, 1\}$; $k_{1,2} = -1/2$, $k_3 = \{-1, 0\}$; $k_{1,2} = 0$, $k_3 = -1$ do not vanish at $t \to \pm\infty$, while the remaining 4 subfamilies, $\{k_{1,2} = -1, k_3 = \{-1/2, 0, 1/2\}\}$ and $k_{1,2,3} = -1/2$, suggest bell-shaped asymmetric pulses vanishing at infinity. The limits $U(t = -\infty) = U(y \to a_0 + 0)$ and $U(t = +\infty) = U(y \to +\infty)$ are listed in Table 6.

We conclude by noting that other complex-valued transformations leading to real physical field configurations can be suggested. For instance, one may apply the transformation $z(t) = \sqrt{a} \exp(i\Delta t)$. With the parameters chosen as

$$U_0^* = i^{-1-2k_3} U_0 / \Delta, \quad \delta_1 = -i, \quad \delta_2 = \delta_3 = 0, \quad (44)$$

the classes $k_{1,2,3} = \{-1/2, -1/2, -1/2\}$ and $k_{1,2,3} = \{0, -1, -1\}$ then result in periodic amplitude modulations. The corresponding field configurations (see Fig. 12) are given as

$$U(t) = \frac{U_0}{(1 + a - 2\sqrt{a} \cos(\Delta t))^{-k_3}}, \quad \delta_t(t) = \Delta. \quad (45)$$

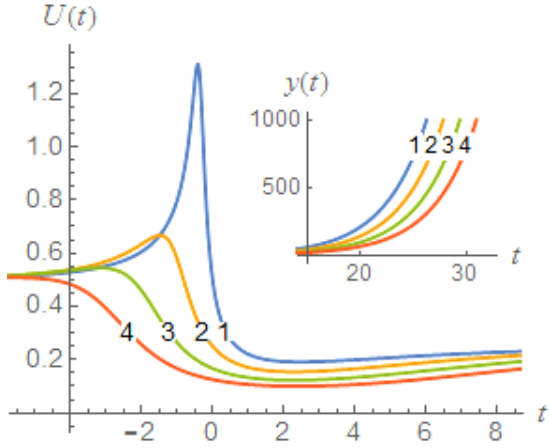
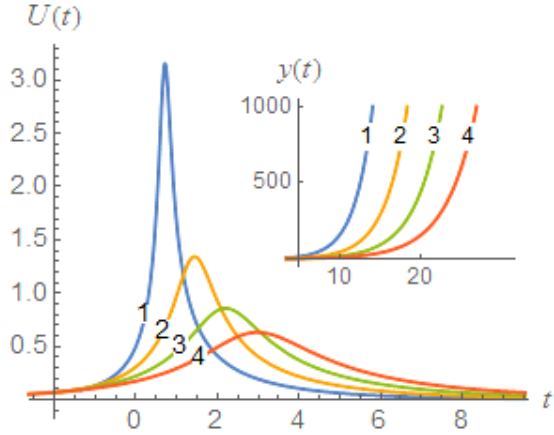

Fig. 10. Class $k_{1,2,3} = \{0, 0, -1\}$. Constant detuning case, $\Delta = 1$. Pulse shapes for $U_0^* = 1$, $a_0 = -2$, $\delta_1 = 1, 1-i/2, 1-i, 1-3i/2$ (curves 1,2,3,4), $\delta_2 = \overline{\delta_1}$ and $\delta_3 = 2$.

Fig. 11. Class $k_{1,2,3} = \{-1, -1, 0\}$. Constant detuning case, $\Delta = 1$. Pulse shapes for $U_0^* = i$, $a_0 = -0.2$, $\delta_1 = 1/2 + i$, $\delta_2 = \overline{\delta_1} = 1/2 - i$ and $\delta_3 = 1.2, 1.7, 2.2, 2.7$ (curves 1,2,3,4).

| $k_{1,2,3}$ | $U(t = -\infty)$ | $U(t = +\infty)$ |
|---|---|---|
| $\{-1, -1, -1\}$ | $U_0 / (\lambda_3 (1 + a_0^2))$ | 0 |
| $\{-1, -1, -1/2\}$ | 0 | 0 |
| $\{-1, -1, 0\}$ | 0 | 0 |
| $\{-1, -1, 1/2\}$ | 0 | 0 |
| $\{-1, -1, 1\}$ | 0 | $U_0 / (2\lambda_1 + \lambda_3)$ |
| $\{-1/2, -1/2, -1\}$ | $U_0 / (\lambda_3 \sqrt{1 + a_0^2})$ | 0 |
| $\{-1/2, -1/2, -1/2\}$ | 0 | 0 |
| $\{-1/2, -1/2, 0\}$ | 0 | $-U_0 / (2\lambda_1 + \lambda_3)$ |
| $\{0, 0, -1\}$ | $U_0 / \lambda_3$ | $U_0 / (2\lambda_1 + \lambda_3)$ |

Table 6. The limits $U(t = \pm\infty)$ for nine classes producing real pulses by the complex-valued transformation $z = (1 + i y(t))/2$. It is supposed that $\lambda_3 > 0$ and $2\lambda_1 + \lambda_3 > 0$.

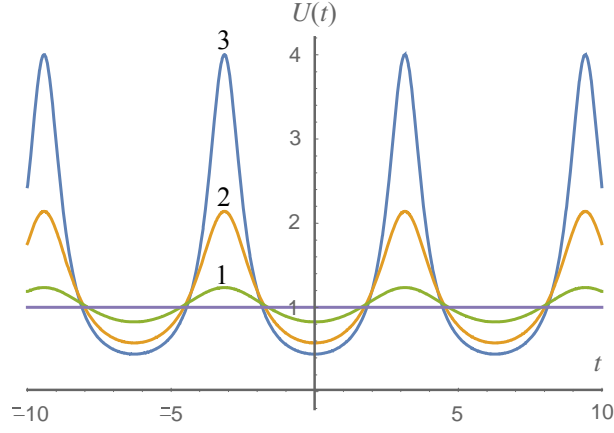

Fig. 12. Class $k_{1,2,3} = \{0, -1, -1\}$. Constant-detuning case, $\delta_t(t) = \Delta = \mathrm{const}$. Periodic amplitude modulation generated by the complex-valued transformation $z(t) = \sqrt{a}\exp(i\Delta t)$: $U_0^* = iU_0/\Delta$, $\delta_1 = -i$, $\delta_2 = \delta_3 = 0$, $\Delta = 1$, $U_0 = 1$, $a = 0.01, 0.1, 0.25$ (curves 1,2,3).

Interestingly, with the same transformation $z(t) = \sqrt{a}\exp(i\Delta t)$ and with parameters

$$U_0^* = -iU_0/\Delta,\quad \delta_1 = -i\Delta_1/\Delta,\quad \delta_2 = -\delta_3 = i\Delta_2/\Delta, \qquad (46)$$

another class, $k_{1,2,3} = \{-1, 0, 0\}$, presents a *constant-amplitude* field configuration with periodic modulation of the detuning:

$$U(t) = U_0,\quad \delta_t(t) = \Delta_1 + \frac{(1-a)\Delta_2}{1 + a - 2\sqrt{a}\cos(\Delta t)}. \qquad (47)$$

For different values of input parameters $a, \Delta_{1,2}$ we have periodic level-crossing, level-glancing or non-crossing detuning modulations (Fig. 13).

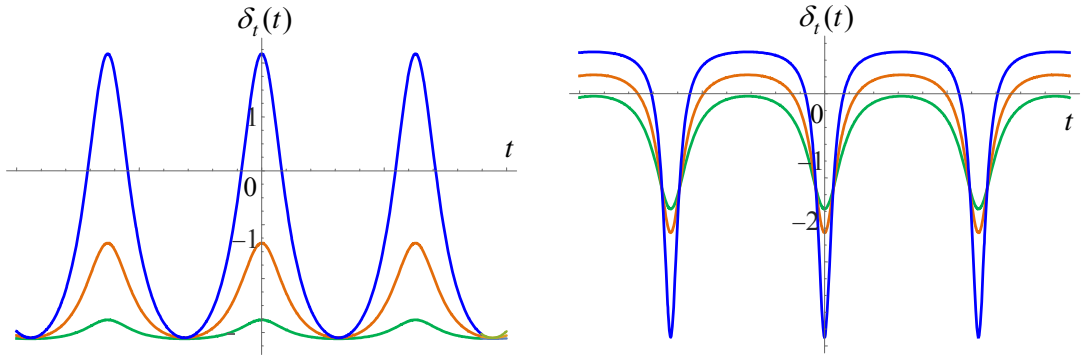

Fig. 13. Class $k_{1,2,3} = \{-1, 0, 0\}$. Constant-amplitude case, $U(t) = U_0 = \mathrm{const}$. Periodic detuning modulations generated by the complex-valued transformation $z(t) = \sqrt{a}\exp(i\Delta t)$.

## 9. Discussion

Thus, we have derived 35 five-parametric classes of two-state models solvable in terms of the general Heun function. The classes are defined by two generating functions which are referred to as the amplitude- and detuning-modulation functions. The actual field configurations, that is, the Rabi frequency and the detuning, are then generated applying a real or complex-valued transformation of the independent variable. Many of the derived classes present generalizations of the six known 3-parametric hypergeometric classes for which the solution of the two-state problem is written in terms of the Gauss hypergeometric function. In several cases the generalization is achieved by multiplying the amplitude modulation function of the corresponding prototype hypergeometric class by an extra factor including an additional parameter. In all these cases as well as in the cases when the amplitude modulation function is not modified compared with the hypergeometric prototype an additional generalization comes due to an extra term in the detuning modulation function. Finally, many classes suggest amplitude modulation functions not discussed before.

The detuning modulation function is the same for all the derived 35 classes. This function involves four arbitrary parameters, that is, two more than the hypergeometric classes. These parameters in general are complex and should be chosen so that the resultant detuning is real for the applied complex-valued transformation of the independent variable. The generalization of the detuning modulation function to the four-parametric case is the most notable extension since many useful properties of the two-state models described by the Heun equation are due to namely the additional parameters involved in this function.

We note that the technique applied for derivation of the presented solvable classes essentially rests on the property (3) [52-55] of the solvable cases of the time-dependent Schrödinger equations (1). This approach assumes two successive steps. The first step consists in identification of the basic solvable models, as listed in above Tables 1-5, by means of transformation of the dependent variable. The transformation of the independent variable is explicitly used only in the second step when the actual field configurations are specified, see Eqs. (11),(12). Unlike this approach, in the most of the cases discussed in literature the stress is done, perhaps by historical reasons, on the transformation of the independent variable. However, a short examination shows that this is a rather restrictive approach which is potent to produce only a very few results. Indeed, suppose $\varphi = 1$, i.e., $\alpha_{1,2,3} = 0$, so that the Heun function appears in the final solution (13) without a pre-factor. It then follows from Eqs. (14) that in this case should be $k_{1,2,3} \neq -1$. This leads to only 4 classes out of presented 35: $k_{1,2,3} = \{-1/2,-1/2,0\}$, $\{-1/2,0,-1/2\}$, $\{0,-1/2,-1/2\}$ and $\{-1/2,-1/2,-1/2\}$. Thus, the conclusion is that the approach based on the property (3) of solvable models is significantly more advanced one. The four mentioned cases with $\varphi = 1$ are indicated in Fig.1 by three triangles and a square which indicates the last class. Note that the first of these four classes presents a generalization of a class solvable in terms of the Gauss hypergeometric functions [53-55] to a more general type of frequency modulation due to the term proportional to $\delta_3$ in Eq. (12). Furthermore, the next two classes indicated by triangles are modifications of the first class for the singular points $\{0,a\}$ and $\{1,a\}$, respectively. The most interesting is the case $k_{1,2,3} = \{-1/2,-1/2,-1/2\}$, which suggests generalizations of both detuning- and amplitude-modulation functions as compared with its hypergeometric prototype. This case was treated in [75,76], however, the treatment there is restricted to the case $\delta_3 = 0$, which, of course, notably weakens the extent of the generalization.

We have determined the parameters of the general Heun function involved in the final solution of the initial two-state problem and have presented the power series expansion of this function. Furthermore, we have mentioned a particular series expansion of the general Heun function in terms of the Gauss hypergeometric functions that is convenient for derivation of particular closed form finite sum solutions.

Discussing the constant detuning case, we have presented several particular families of pulses generated by real transformation of the independent variable. These families include both symmetric and asymmetric members. Among all the families only 10 provide finite area pulses, that is, pulses that vanish at infinity. The members of these families are in general symmetric or asymmetric two-peak pulses with controllable distance between the peaks and controllable amplitude of each of the peaks. We have shown that the edge shapes, the distance between the peaks as well as the amplitude of the peaks are controlled almost independently, by different parameters. We have identified the parameters controlling each of the mentioned features and have discussed other basic properties of pulse shapes.

We have shown that the pulse edges may become step-wise functions and determined the positions of the limiting vertical-wall edges. We have shown that the pulse width is then controlled by only two of the involved parameters. For some values of these parameters the pulse width diverges and for some other values the pulses become infinitely narrow. We have shown that the effect of the two mentioned parameters is almost similar, that is, both parameters are able to independently produce pulses with almost the same shape and width.

We have determined the conditions for generation of pulses of almost indistinguishable shape and width, and have presented several such examples.

The derived classes provide a rich variety of field configurations generated by different real or complex choices of the independent variable transformation. For instance, apart from the above families of constant detuning pulses obtained by real $z(t)$, many other such constant detuning families are suggested by a complex transformation of the form $z = (1 + i\, y(t))/2$ or $z(t) = \sqrt{a}\, \exp(i\Delta t)$. Examples with variable detuning include numerous models with constant amplitude of the field, several configurations with periodic modulation of the amplitude, a large variety of families of chirped pulses, level-glancing models and models describing double or periodic level-crossings. Notably, the classes also suggest models describing excitation of a two-level atom by bi-chromatic laser fields. This is not the whole list, there are several other possibilities.

We would like to conclude by a brief note concerning possible applications of above models. First, we mention the appreciable richness of the set of the Heun models as compared with the hypergeometric set (the number of the Heun models prevails by an order of the magnitude). Second, the Heun models suggest a variety of distinct features not present in the hypergeometric ancestors (e.g., double- and multiple level-crossings, two-peak pulses, etc.). For this reason, one may foresee many applications in quantum and atom optics, e.g., in quantum engineering via manipulation of quantum structures with prescribed properties [94,95] by laser pulses, in the theory of chemical reactions including cold atom association in quantum gases, etc., however, it is difficult if not impossible to list all possibilities.

We would like to mention here just a useful point of mathematical character. In several cases the application of the presented Heun models may be advantageous for general considerations (this may be the case even if the qualitative behavior of the pulses is close to those discussed using hypergeometric models). This is because in several cases the models allow closed form solutions based on series expansions of the involved Heun functions. An example supporting this observation is the solution of the two-state problem for the constant-amplitude field configuration (47) describing a periodic level-crossing process. It is readily checked using Eqs. (14)-(17) that the parameters of the general Heun function involved in the final solution (18) all are real and such that allow a series expansion in terms of the incomplete Beta-functions [96]. For a certain infinite (countable) set of parameters this series is terminated thus resulting in closed-form finite-sum solutions. We intend to discuss the applications of these solutions in a future research.

**Acknowledgments**

This research has been conducted within the scope of the International Associated Laboratory (CNRS-France & SCS-Armenia) IRMAS. The research has received funding from the European Union Seventh Framework Programme (FP7/2007-2013) under grant agreement No. 295025 – IPERA. The work has been supported by the Armenian State Committee of Science (SCS Grant No. 13RB-052).

**References**

1. L.D. Landau, Phys. Z. Sowjetunion **2**, 46 (1932).
2. C. Zener, Proc. R. Soc. London, Ser. A **137**, 696 (1932).
3. E. Majorana, Nuovo Cimento **9**, 43 (1932).
4. E.C.G. Stückelberg, Helv. Phys. Acta. **5**, 369 (1932).
5. M.S. Child, *Molecular Collision Theory* (Academic Press, London, 1974).
6. E.E. Nikitin and S.Ya. Umanski, *Theory of Slow Atomic Collisions* (Springer-Verlag, Berlin, 1984).
7. H. Nakamura, *Nonadiabatic Transition: Concepts, Basic Theories and Applications* (World Scientific, Singapore, 2012).
8. B.W. Shore, *The Theory of Coherent Atomic Excitation* (Wiley, New York, 1990).
9. B.W. Shore, *Manipulating Quantum Structures Using Laser Pulses* (Cambridge University Press, New York, 2011).
10. H.J. Metcalf and P. van der Straten, *Laser Cooling and Trapping* (Springer-Verlag, New York, 1999).
11. A.P. Kazantsev, G.I. Surdutovich, and V.P. Yakovlev, *Mechanical Action of Light on Atoms* (World Scientific, Singapore, 1990).
12. Pierre Meystre, *Atom Optics* (Springer Verlag, New York, 2001).
13. H. Nakamura, Int. Rev. Phys. Chem. **10**, 123 (1991).
14. H. Nakamura, Ann. Rev. Phys. Chem. **48**, 299 (1997).
15. C. Zhu, Y. Teranishi and H. Nakamura, Adv. Chem. Phys. **117**, 127 (2001).
16. *Electron Transfer in Inorganic, Organic, and Biological Systems*, Advances in Chemistry Series **228**, eds. J. Bolton, N. Mataga and G. Mclendon (American Chemical Society, Washington, D.C., 1991).
17. D. DeVault, *Quantum Mechanical Tunnelling in Biological Systems* (Cambridge University Press, Cambridge, 1984).
18. D.E. Shaw et al., Science **330**, 341 (2010).
19. W.H. Zurek, U. Dorner, and P. Zoller, Phys. Rev. Lett. **95**, 105701 (2005).
20. B. Damski, Phys. Rev. Lett. **95**, 035701 (2005).
21. R. Barankov and A. Polkovnikov, Phys. Rev. Lett. **101**, 076801 (2008).
22. J. Dziarmaga, Adv. Phys. **59**, 1063 (2010).
23. F. Gaitan, Phys. Rev. A **68**, 052314 (2003).
24. D.M. Berns, W.D. Oliver, S.O. Valenzuela, A.V. Shytov, K.K. Berggren, L.S. Levitov, and T.P. Orlando, Phys. Rev. Lett. **97**, 150502 (2006).
25. K. Smith-Mannschott, M. Chuchem, M. Hiller, T. Kottos, and D. Cohen, Phys. Rev. Lett. **102**, 230401 (2009).
26. A.M. Ishkhanyan, Eur. Phys. Lett. **90**, 30007 (2010).
27. M. Jona-Lasinio, O. Morsch, M. Cristiani, N. Malossi, J.H. Muller, E. Courtade, M. Anderlini, and E. Arimondo, Phys. Rev. Lett. **91**, 230406 (2003).
28. M.-O. Mewes, M.R. Andrews, D.M. Kurn, D.S. Durfee, C.G. Townsend, and W. Ketterle, Phys. Rev. Lett. **78**, 582 (1997).
29. N.V. Vitanov and K.-A. Suominen, Phys. Rev. A **56**, R4377 (1997).

30. A. Ishkhanyan, Phys. Rev. A **81**, 055601 (2010).
31. I. Tikhonenkov, E. Pazy, Y.B. Band, M. Fleischhauer, and A. Vardi, Phys. Rev. A **73**, 043605 (2006).
32. A.M. Ishkhanyan, B. Joulakian and K.-A. Suominen, Eur. Phys. J. D **48**, 397 (2008).
33. D. Sun, A. Abanov and V.L. Pokrovsky, Eur. Phys. Lett. **83**, 16003 (2008).
34. A. Ishkhanyan, B. Joulakian, and K.-A. Suominen, J. Phys. B **42**, 221002 (2009).
35. F.R. Braakman, P. Barthelemy, C. Reichl, W. Wegscheider and L.M.K. Vandersypen, Nature Nanotechnology **8**, 432 (2013).
36. W. Wernsdorfer and R. Sessoli, Science **284**, 133 (1999).
37. A.F. Terzis, E. Paspalakis, J. Appl. Phys. **97**, 023523 (2005).
38. P. Földi, M.G. Benedict, F.M. Peeters, Phys. Rev. A **77**, 013406 (2008).
39. S.J. Parke, Phys. Rev. Lett. **57**, 1275 (1986).
40. W.C. Haxton, Phys. Rev. D **35**, 2352 (1987).
41. M. Blennow and A.Yu. Smirnov, Adv. High Energy Phys. **2013**, 972485 (2013).
42. E. E. Nikitin, Opt. Spectrosc. **6**, 431 (1962)
43. E. E. Nikitin, Discuss. Faraday Soc. **33**, 14 (1962).
44. E.E. Nikitin, Annu. Rev. Phys. Chem. **50**, 1-21 (1999).
45. N. Rosen and C. Zener. Phys. Rev. **40**, 502 (1932).
46. Yu.N. Demkov and M. Kunike, Vestn. Leningr. Univ. Fis. Khim. **16**, 39 (1969).
47. K.-A. Suominen and B.M. Garraway, Phys. Rev. A **45**, 374 (1992).
48. A. Bambini and P.R. Berman, Phys. Rev. A **23**, 2496 (1981).
49. F.T. Hioe and C.E. Carroll, Phys. Rev. A **32**, 1541 (1985).
50. F.T. Hioe and C.E. Carroll, J. Opt. Soc. Am. B **3**(2), 497 (1985).
51. C.E. Carroll and F.T. Hioe, J. Phys. A **19**, 3579 (1986).
52. A.M. Ishkhanyan, J. Phys. A **30**, 1203 (1997).
53. A.M. Ishkhanyan, Optics Communications **176**, 155 (2000).
54. A.M. Ishkhanyan, J. Contemp. Physics (Armenian Ac. Sci.) **31**(4), 10 (1996).
55. A.M. Ishkhanyan, J. Phys. A **33**, 5539 (2000).
56. C.E. Carroll and F.T. Hioe, J. Opt. Soc. Am. B 5, 1335 (1988).
57. C.E. Carroll and F.T. Hioe, J. Phys. B **22**, 2633 (1989).
58. C.E. Carroll and F.T. Hioe, J. Phys. A **19**, 1151 (1986).
59. C.E. Carroll and F.T. Hioe, Phys. Rev. A **36**, 724 (1987).
60. C.E. Carroll and F.T. Hioe, Phys. Rev. A **42**, 1522 (1990);
61. A.M. Ishkhanyan, J. Phys. A **33**, 5041 (2000).
62. A.M. Ishkhanyan and K.-A. Suominen, Phys. Rev. A **65**, 051403(R) (2002).
63. T.A. Laine and S. Stenholm, Phys. Rev. A **53**, 2501 (1996).
64. N.V. Vitanov and S. Stenholm, Phys. Rev. A **55**, 648 (1997).
65. A.M. Ishkhanyan, Reports (Armenian Ac. Sci.) **102**(4), 320 (2002).
66. A.M. Ishkhanyan and A.M. Manukyan, J. Contemp. Physics (Armenian Ac. Sci.), **37**(4), 1 (2002).
67. A. Erdélyi, W. Magnus, F. Oberhettinger, and F.G. Tricomi, *Higher Transcendental Functions*, vol. **3** (McGraw-Hill, New York, 1955).
68. L.J. Slater, *Generalized hypergeometric functions* (Cambridge University Press, Cambridge, 1966).
69. K. Heun, Math. Ann. **33**, 161 (1889).
70. A. Ronveaux, *Heun's Differential Equations* (Oxford University Press, London, 1995).
71. S.Yu. Slavyanov and W. Lay, *Special functions* (Oxford University Press, Oxford, 2000).
72. F.W.J. Olver, D.W. Lozier, R.F. Boisvert, and C.W. Clark (eds.), *NIST Handbook of Mathematical Functions* (Cambridge University Press, New York, 2010), http://dlmf.nist.gov/31.12.

73. M. Hortacsu, arXiv:1101.0471 [math-ph].
74. A.M. Ishkhanyan and A.E. Grigoryan, J. Phys. A **47**, 465205 (2014).
75. P.K. Jha and Yu.V. Rostovtsev, Phys. Rev. A **82**, 015801 (2010).
76. P.K. Jha and Yu.V. Rostovtsev, Phys. Rev. A **81**, 033827 (2010).
77. N. Svartholm, Math. Ann. **116**, 413 (1939).
78. A. Erdélyi, Duke Math. J. **9**, 48 (1942).
79. A. Erdélyi, Q. J. Math. (Oxford) **15**, 62 (1944).
80. D. Schmidt, J. Reine Angew. Math. **309**, 127 (1979).
81. E.G. Kalnins and W. Miller, SIAM J. Math. Anal. **22**, 1450 (1991).
82. R.S. Sokhoyan, D.Yu. Melikdzanian and A.M. Ishkhanyan, J. Contemp. Physics (Armenian Ac. Sci.), **40**(6), 1 (2005).
83. T.A. Ishkhanyan, T.A. Shahverdyan, A.M. Ishkhanyan, arXiv:1403.7863 [math.CA] (2014).
84. T.A. Ishkhanyan and A.M. Ishkhanyan, AIP Advances **4**, 087132 (2014).
85. E.S. Cheb-Terrab, J. Phys. A **37**, 9923 (2004).
86. A. Ishkhanyan, J. Phys. A **38**, L491 (2005).
87. A. Ishkhanyan, J. Phys. A **34**, L591 (2001).
88. A. Ishkhanyan, K.-A. Suominen, J. Phys. A **36**, L81 (2003).
89. P.P. Fiziev, J. Phys. A **43**, 035203 (2010).
90. M.N. Hounkonnou, A. Ronveaux, Appl. Math. Comput. **209**, 421 (2009).
91. V.A. Shahnazaryan, T.A. Ishkhanyan, T.A. Shahverdyan, and A.M. Ishkhanyan, Armenian Journal of Physics **5**(3), 146 (2012).
92. J.H. Lambert, Acta Helvitica **3**, 128–168 (1758).
93. L. Euler, Acta Acad. Scient. Petropol. **2**, 29 (1783).
94. A.M. Ishkhanyan, Phys. Rev. A **61**, 063611 (2000).
95. A.M. Ishkhanyan, Laser Physics **7**, no.6, 1225 (1997).
96. A.M. Manukyan et al., IJDEA **13**, 231 (2014).